\documentclass[a4paper,reqno,12pt,draft]{article}
\usepackage{amssymb,euscript,bbold}

\newcommand{\be}{\begin{equation}}
\newcommand{\ee}{\end{equation}}
\newcommand{\ba}{\begin{eqnarray}}
\newcommand{\ea}{\end{eqnarray}}
\newcommand{\nl}{\newline}
\newcommand{\ft}{\footnote}
\newcommand{\pr}{\prime}

\begin{document}

\begin{flushright}
QMW-PH-98-42
\end{flushright}
\begin{center}
\Large{\sc M theory, Joyce Orbifolds and Super
Yang-Mills.}\\
\bigskip
{\sc B.S. Acharya}\ft{e-mail:r.acharya@qmw.ac.uk\nl
Work supported by PPARC}\\
\smallskip\large
{\sf Department of Physics,\\
Queen Mary and Westfield College,\\ Mile End Road,\\ London. E1 4NS.}

\end{center}
\bigskip
\begin{center}
{\bf {\sc Abstract}}
\end{center}
We geometrically engineer d=4 N=1 supersymmetric Yang-Mills
theories by considering
M theory on various Joyce orbifolds. We argue that the superpotential
of these models is generated by fractional membrane instantons. The
relation of this superpotential to membrane anomalies is also
discussed.

\newpage

\Large
\noindent
{\bf {\sf 1. Introduction.}}
\normalsize
\bigskip

In recent years string theory and $M$ theory have provided
us with a deeper insight into the dynamics of (mainly supersymmetric)
gauge theories in various dimensions. Difficult questions
in field theory can turn out to have relatively simple answers when
embedded into a string or $M$ theoretic context.

This powerful approach to the study of gauge theories comes, roughly,
in two related forms: one can study physics localised on
branes by decoupling ``bulk degrees of freedom''
or one can consider the physics localised near certain kinds
of singularities in
spacetime, a
construction which goes by the name of geometric engineering.
Constructions of the former type were initiated in \cite{HW} building
on results of many other authors and
are reviewed in \cite{GK}, whereas geometric engineering was
christened in \cite{KV} a paper which was also motivated
by the results of several authors concerning massless states
appearing at singularities. Many references to these works
can be found in \cite{KV} and \cite{K3}.


$M$ theory compactified on a 7-manifold
with $G_2$ holonomy gives rise to a four dimensional theory
with ${\cal N}$ $=$ $1$ supersymmetry. For a given 7-manifold ${\bf X}$
with $G_2$ holonomy, what is the physics of $M$ theory compactified
on ${\bf X}$ ?  At low energies we can use the eleven dimensional
supergravity approximation to $M$ theory and one finds that
the effective four
dimensional supergravity theory describing the
{\it massless} modes,
as first noted in \cite{paptown}, is ${\cal N}$ $=$ $1$ supergravity
coupled to $b_2$ $U(1)$ vector multiplets and $b_3$ neutral chiral
multiplets, where $b_i$ are Betti numbers of
${\bf X}$.
Such a supergravity theory is relatively uninteresting physically.
However, this particular massless spectrum, is that
of $M$ theory compactified on a {\it smooth} 7-manifold of $G_2$
holonomy. On the other hand, an important
lesson learned in string theory and $M$ theory
over the past few years has been that interesting physics can arise
when singularities in a ``compactification manifold'' appear. 
The main purpose of this paper is to initiate a study of the physics
associated with singularities in
spaces with $G_2$ holonomy. In particular, we will 
construct and study `local models'
of natural (and perhaps the simplest) types of singularities in this
context; we will use these models to geometrically engineer
four dimensional ${\cal N}$ $=$ $1$ supersymmetric
Yang-Mills theories. Geometric engineering of ${\cal N}$ $=$
$1$ theories has been studied in detail previously
in the context of
Calabi-Yau backgrounds \cite{n1}.

To motivate the types of singularities we will study consider the
following. Let ${\bf S}$ be a ${\bf K3}$ surface and 
$\{{{\omega}_i}\}$ a triplet of Kahler forms defining a hyperkahler
structure on ${\bf S}$. Let ${\bf T^3}$ be a flat 3-torus and
$dx_i$ a basis for its cotangent bundle. Then the product manifold
${\bf {S}}{\times}{\bf T^3}$ admits a parallel $G_2$ structure defined as
follows.
\ba
{\varphi} &=& {\delta}^{ij}{\omega}_i {\wedge}{dx_j} +
{1 \over 6}{\epsilon}^{ijk}{dx_i}{\wedge}{dx_j}{\wedge}{dx_k}\\
\star{\varphi} &=& {1 \over 6}{\delta}^{ij}{\omega}_i {\wedge}{\omega}_j
+ {1 \over 2}{\epsilon}^{ijk}{\omega}_i {\wedge}{dx_j}{\wedge}{dx_k}
\ea

In the above ${\delta}^{ij}$ is the identity matrix, ${\epsilon}^{123}$
$=$ $1$ and is totally antisymmetric and we have used the Einstein
summation convention. $SO(3)$ acts naturally on
both triplets ${\omega}_i$ and $dx_i$. The above $G_2$ structure is
hence manifestly $SO(3)$-invariant.

Let ${\bf K}$ denote a finite subgroup of $SO(3)$ under which
the triplets ${\omega}_i$ and $dx_i$ transform in the {\it same} 
representation. Then the orbifold ${\bf {O_K}}$ $\equiv$
${\bf (S{\times}{T^3})/{K}}$
inherits the above parallel $G_2$ structure. Furthermore, if ${\bf K}$ is
{\it not} a subgroup of $SO(2)$ $\subset$ $SO(3)$, then
${\bf {O_K}}$ will {\it not} inherit a parallel $SU(3)$ structure.  
We will always choose ${\bf K}$ $\not\subset$ $SO(2)$ in what follows.
This technical restriction ensures ${\cal N}$ $=$ $1$ supersymmetry
and not ${\cal N}$ $=$ $2$.

If we now consider $M$ theory on ${\bf {O_K}{\times}{R^4}}$, then we
obtain an effective four dimensional theory with ${\cal N}$ $=$ $1$
supersymmetry (assuming as we will do that such an orbifold
compactification is well defined in $M$ theory). There are now
essentially four cases to consider. $(i)$ ${\bf K}$ acts freely on
${\bf {S}{\times}{T^3}}$ {\it and} ${\bf S}$ is a non-singular
${\bf K3}$ surface; $(ii)$ ${\bf K}$ acts freely on 
${\bf {S}{\times}{T^3}}$ {\it and} ${\bf S}$ is a singular
${\bf K3}$ surface; $(iii)$ ${\bf K}$ does not act freely
and ${\bf S}$ is non-singular; $(iv)$ ${\bf K}$ does not
act freely and ${\bf S}$ is singular. By singular, we mean
a ${\bf K3}$ surface with {\sf A-D-E} orbifold singularities,
so that the notion of hyperkahler structure still makes sense.
In case $(i)$ 
${\bf {O_K}}$ is a smooth manifold and therefore the low
energy physics is that we described above. ${\bf {O_K}}$ clearly
has singularities in the remaining three cases.
In this paper we will describe the physics localised
near singularities of type $(ii)$. We hope to address types
$(iii)$ and $(iv)$ in future investigations.

For cases of type $(ii)$ the space ${\bf O_K}$ can naturally
be thought of as being fibered by the ${\bf K3}$ surfaces ${\bf S}$.
${\bf K3}$ fibrations of Joyce orbifolds \cite{J1} 
with $G_2$ holonomy
were discussed
in the context of $M$ theory/heterotic duality in \cite{bsa1} and
have been further investigated in \cite{fibers}.
 
We are interested ultimately in $M$ theory
physics near the singularities of ${\bf O_K}$. Since in the case
of type $(ii)$ these correspond locally to {\sf A-D-E} surface
singularities, in order to construct a local model we will `replace'
${\bf S}$ with an orbifold ${\bf {C^2}/{G}}$, where ${\bf G}$
is a finite subgroup of $SU(2)$. In other words we
will be considering orbifolds of the form 
${\bf J^G_K}$ $\equiv$ ${\bf ({C^2}/{G}{\times}{T^3})/{K}}$.
By abuse of notation, if we let ${\omega}_i$ denote the
hyperkahler structure on ${\bf {C^2}/{G}}$, then equations
$(1)$ and $(2)$ define a parallel $G_2$-structure on ${\bf J}$.

We can further divide the orbifolds ${\bf J}$ into two types.
Since ${\bf K}$ acts freely on ${\bf {C^2}/{G}{\times}{T^3}}$,
these differ according to whether or
not ${\bf K}$ acts freely on ${\bf T^3}$. We will only
consider
cases in which ${\bf K}$ does act freely on ${\bf T^3}$, 
in which case
${\bf J}$ is naturally endowed with the structure of
a fibration whose fibers are the ${\bf {C^2}/{G}}$ orbifolds
and whose base is the smooth 3-manifold
${\bf {T^3}/{K}}$. The singular set of the orbifold is
${\bf M_K}$ $\equiv$
${\bf (\{0\}{\times}{T^3})/{K}}$ (where ${\bf \{0\}}$ is the
origin in ${\bf {C^2}/{G}}$) and we can see
that the singular set of ${\bf J}$ consists of a
family of {\sf A-D-E} singularities fibered over
${\bf {T^3}/{K}}$. 

In the next section we will study families of {\sf A-D-E}
singularities and compute the four dimensional massless spectrum
localised near these singularities. The results of this section
are in a similar spirit to results which have appeared in
\cite{morple,asp,KV} concerning families of {\sf A-D-E} singularities
in Calabi-Yau spaces.
We then describe 
some explicit examples of the orbifolds ${\bf J^G_K}$ for which
${\bf G}$ is any finite {\sf A-D-E} subgroup of $SU(2)$ and
${\bf K}$ is isomorphic to ${\bf {Z_2}{\times}{Z_2}}$ or
${\bf {Z_2}{\ltimes}{Z_4}}$.
It turns out that the geometrically engineered models
obtained from $M$ theory on ${\bf J^G_K}$ are described
at low energies by pure ${\cal N}$ $=$ $1$ super Yang-Mills
theory with {\sf A-D-E} gauge group corresponding to
${\bf G}$.
 
We then take a slight detour and propose a Type IIA dual
description of some of the $M$ theory models involving
D6-branes. This section also clarifies certain aspects of
the computation of the spectrum in section two. Following
a brief review of ${\cal N}$ $=$ $1$ super Yang-Mills theory
(which is included mainly for completeness) we discuss 
the ${1 \over 2}$ supersymmetric BPS states present in
the geometrically engineered models. We show firstly
that the models engineered using ${\bf J^G_K}$ contain
fractional $M$2-brane instantons (of a type discussed in
\cite{frac}), 
which correspond in field theory
to states with fractional instanton numbers; secondly we
argue that fractional instantons of a particular topological
charge can generate a superpotential.
This superpotential
matches precisely that expected from field theory considerations.
In \cite{brodie} it was argued that fractional instantons generate
the superpotential 
in an alternative construction \cite{MQCD} (now known as
$M$ QCD) of the
same field theory (with gauge group $SU(n)$)
using Type IIA/$M$ branes. We refer the reader to \cite{GK}
for further references to ${\cal N}$ $=$ $1$ models
constructed using branes. 
Furthermore,
it has been argued in \cite{gom1} in field theory and in
\cite{gom2} in the geometrically engineered models of
\cite{n1}
that such a superpotential
can be generated by torons in a certain limit.
Several other aspects of fractional instantons in field theory
have also been discussed in \cite{zab}. 

In the final section of this paper we describe a relation
between the instanton generated superpotential and
the parity anomaly on the $M$2-brane world-volume. The
result here is that, modulo an assumption, $M$2-branes
indeed know about the anomaly seen in the field theory.
This is a satisfying confirmation of the main results
which
complement nicely the results of \cite{Wit} in which
superpotentials were generated using $M$5-brane instantons
in Calabi-Yau spaces. 

{\sf Note Added.} After completing the first version of this paper
we were informed by G. Moore of the forthcoming paper \cite{harvmoore}
in which there is overlap with the present work.

\bigskip
\Large
\noindent
{\bf {\sf 2. Supersymmetric Families of A-D-E singularities.}}
\normalsize
\bigskip

The orbifolds ${\bf J^G_K}$ can be thought of as families
of ${\bf {C^2}/{G}}$ orbifolds parameterised by ${\bf M_K}$.
${\bf M_K}$ is naturally a supersymmetric 3-cycle. This is 
straightforward to establish since ${\bf \{0\}{\times}{T^3}}$
is a supersymmetric cycle in ${\bf {C^2}/{G}{\times}{T^3}}$ and
${\bf K}$ preserves its volume form. Therefore we are considering
the orbifold singularities of ${\bf {J^G_K}}$ as local models
for supersymmetric (ie associative) families of {\sf A-D-E}
surface singularities which might develop in
spaces of $G_2$ holonomy as we move through the moduli space of
metrics. 

What is the $M$ theory physics localised near the singularities
of ${\bf J^G_K}$ ? In order to answer this question, we should first 
compute what massless modes are supported near the singularities.
We will actually give a more general answer which computes the massless
spectrum localised near any 3-dimensional supersymmetric family of
{\sf A-D-E} singularities. 
Therefore we consider a manifold
${\bf X}$ with $G_2$ holonomy. As we vary the $G_2$ holonomy metrics
on ${\bf X}$ we assume that we
encounter a 3-dimensional family of {\sf A-D-E}
singularities parametrised by a 3-manifold ${\bf M}$ such that
${\bf M}$ is a supersymmetric submanifold of the singular limit
of ${\bf X}$.

Begin first by considering $M$ theory near an {\sf A-D-E} singularity
of the form
${\bf \{x\}{\times}{R^7}}$, where ${\bf \{x\}}$ is the singular
point in a 4-space with hyperkahler structure.
The physics localised near such a
singularity of spacetime is described by
7-dimensional super Yang-Mills theory on ${\bf R^7}$
with {\sf A-D-E} gauge group $H$
(determined by which {\sf A-D-E} subgroup of $SU(2)$ ${\bf G}$ is).
The singularities of spacetime we wish to consider are locally of the
form ${\bf \{x\}{\times}{M}{\times}{R^4}}$ which occur inside
a spacetime ${\bf {X}{\times}{R^4}}$ with ${\bf X}$ a space of
$G_2$ holonomy. 
From the point of view of the 7-dimensional Yang-Mills theory
we have essentially replaced ${\bf R^7}$ with ${\bf {M}{\times}{R^4}}$.
We should thus consider
7-dimensional super Yang-Mills theory on ${\bf {M}{\times}{R^4}}$
and compute the massless modes in ${\bf R^4}$ {\it a la Kaluza-Klein}.

Since generically
${\bf M}$ will not admit any parallel spinors one might first
think that all supersymmetry is broken by this compactification.
However, ${\bf M}$ is a supersymmetric 3-cycle in a space with
$G_2$ holonomy and this means that the Yang-Mills theory is
twisted and is indeed supersymmetric. This is very much in the spirit
of discussions in \cite{morple,dtop}.
We will illustrate this twisting shortly.

In ${\bf R^7}$ the symmetries of super Yang-Mills theory include
\newline
$Spin(3){\times}Spin(6,1)$. $Spin(3)$ is the {\sf R}-symmetry group
and $Spin(6,1)$ the cover of the Lorentz group. $Spin(3)$ also
acts as rotations of the hyperkahler structure associated with
the {\sf A-D-E} singularity.
The fields of the theory
are all massless and consist of three scalars ${\phi}_i$ transforming
in the ${\bf (3,1)}$, gauge bosons transforming in the ${\bf (1,7)}$
and 16 fermions in the ${\bf (2,8)}$. When we `compactify' to
${\bf {M}{\times}{R^4}}$ the $Spin(6,1)$ symmetry is broken to
\newline
$Spin(3)^{\pr}{\times}{SL(2,{\bf C})}$. Here $Spin(3)^{\pr}$
is the cover of the structure group of ${\bf M}$ and
$SL(2,{\bf C})$ is the cover of the Lorentz group in 4-dimensions.
However, as is perhaps evident in equations $(1)$ and $(2)$,
the fact that ${\bf M}$ is a supersymmetric submanifold inside
a space with $G_2$ holonomy means that the theory is `twisted':
in fact the two $Spin(3)$ groups are identified under this twisting.
In this `twisted' theory the symmetry group is just
$SO(3){\times}SL(2,{\bf C})$ with $SO(3)$ the structure group
of ${\bf M}$. We give an alternative and clearer
description of this twisting in terms of wrapped D6-branes
in Type IIA theory later in this paper.

The fields and their transformation properties under
this twisted symmetry group are, in the bosonic sector 
two 1-forms on ${\bf M}$
which are scalars in ${\bf R^4}$. These transform as ${\bf 2(3,1)}$.
There are also scalars on ${\bf M}$ which are gauge fields on
${\bf R^4}$; these transform as ${\bf (1,2{\otimes}{\bar 2})}$.
The fermions of the theory are organised in the following fashion.
There are two spacetime fermions one of which is a scalar on ${\bf M}$
the second of which is a 1-form on
${\bf M}$. The fermions thus transform as ${\bf (1+3,2+{\bar 2})}$.
These fields will be massless if they are zero modes of the relevant
derivative operator on ${\bf M}$. In this case the relevant
operator is just the Laplacian on ${\bf M}$.  

We thus find that the massless
spectrum supported near this family of singularities consists of
$1$ gauge field, $b_1$ scalars and $1+b_1$ fermions.
($b_1$ is the
first Betti number of ${\bf M}$). We should also recall that
in the uncompactified theory all the fields transform in the adjoint
representation of the {\sf A-D-E} gauge group. This however, does
not imply that the gauge group of the four dimensional theory
is the same as that in seven dimensions; as noted in the context
of $F$-theory, when obtaining lower dimensional gauge fields
from higher dimensions by `fibering' {\sf A-D-E} singularities
over some space ${\bf B}$, certain monodromies in the
fibration can act as {\it outer} automorphisms of the {\sf A-D-E}
group and hence break the gauge group to the subgroup commuting
with this automorphism. In this paper we will mainly consider
situations in which this does {\it not} occur. Whether or not
such a monodromy occurs depends upon the choice of the group
${\bf K}$ in the spaces ${\bf J^G_K}$. 
If we assume that such outer automorphisms do not occur then all
the above massless fields transform in the adjoint representation of
the {\sf A-D-E} gauge group.  Otherwise they transform under the
broken gauge group. 
The massless field
content is then precisely that of pure 
${\cal N}$ $=$ $(1+b_1)$ super Yang-Mills theory in
four dimensions. 

For the
orbifolds ${\bf J^G_K}$, ${\bf M}$ is ${\bf {T^3}/{K}}$ and has
$b_1$ $=$ $0$. Hence the massless spectrum consists of 
pure ${\cal N}$ $=$ $1$ super Yang-Mills theory with
{\sf A-D-E} gauge group determined by ${\bf G}$. 

As a check of
this result consider the special cases 
when ${\bf K}$ is the trivial group or a subgroup of $SO(2)$.
When ${\bf K}$ is trivial, ${\bf J^G_{1}}$ $\cong$
${\bf {C^2}/{G}{\times}{T^3}}$ and ${\bf M}$ $\cong$ ${\bf T^3}$.
Hence our computation correctly predicts that the massless modes
supported near the singularity are those of
pure ${\cal N}$ $=$ $4$ super Yang-Mills theory. In the case
when ${\bf K}$ is a subgroup of $SO(2)$ {\it and} ${\bf J^G_K}$
admits a parallel $SU(3)$ structure, ${\bf M}$ $\cong$
${\bf {CP^1}{\times}{S^1}}$ and our computation correctly
predicts a spectrum \cite{morple,asp} 
which matches that of pure ${\cal N}$
$=$ $2$ super Yang-Mills theory.

For concreteness we introduce some particular examples of spaces
with the properties we require, namely a well defined $G_2$
holonomy structure and an associative family of ${\bf {C^2}/{G}}$
hyperkahler orbifold singularities. In fact, for these examples,
${\bf {K}}$ $\cong$ 
${\bf {Z_2}{\times}{Z_2}}$ or ${\bf {Z_2}{\ltimes}{Z_4}}$
and we can define one such space for
each {\sf A-D-E} subgroup of $SU(2)$.

\bigskip
\large
\noindent 
{\bf {\sf 2.1 Examples.}}
\bigskip
\normalsize

Our first example orbifolds ${\bf J^G_{{Z_2}{\times}{Z_2}}}$,
are defined as follows. We begin with the case ${\bf G}$
$\cong$ ${\bf Z_n}$ and
${\bf R^7}$ $\equiv$ ${\bf {C^2}{\times} {R^3}}$.
Let $({z_1},{z_2})$, and $x_i$ ($i$ = $1,2,3$) denote coordinates for the
${\bf C^2}$ and ${\bf R^3}$ factors respectively.
Form ${\bf C^2}{\times}{\bf T^3}$ by identifying:
\be
(x_i ) \equiv (x_i + 1)
\ee
 
Define three finite groups of isometries of the above 
${\bf {C^2}{\times}{T^3}}$, which are respectively isomorphic
to ${\bf Z_n}$, ${\bf Z_2}$ and ${\bf Z_2}$ and which together
generate 
${\bf \Gamma}$ $\cong$ ${\bf ({Z_2}{\times}{Z_2}){\ltimes}{Z_n}}$:
\ba
{\bf Z_n}(\alpha): (z_1,z_2,x_1,x_2,x_3) &=&
({e^{2\pi i/n}}z_1,{e^{-{2\pi i/n}}}z_2,x_1,x_2,x_3)\\\nonumber
{\bf Z_2}(\beta):  (z_1,z_2,x_1,x_2,x_3) &=&
(-z_1,z_2,-x_1 + {1 \over 2},-x_2,x_3 + {1 \over 2})\\
{\bf Z_2}(\gamma): (z_1,z_2,x_1,x_2,x_3) &=&
(-z_1^*,-z_2^*,-x_1,x_2 + {1 \over 2},-x_3)\nonumber
\ea 
We have denoted by $({\alpha},{\beta},{\gamma})$ the generators
of the three cyclic groups and $*$ denotes complex conjugation.

The orbifold ${\bf J^{Z_n}_{{Z_2}{\times}{Z_2}}}$ 
is defined as the
quotient ${\bf ({C^2}{\times}{T^3})/{\Gamma}}$.

\noindent Let ${\bf F}$ be the subgroup of ${\bf \Gamma}$ which
acts with fixed points and denote by ${\bf {\Gamma}^{\prime}}$
the quotient group ${\bf \Gamma}/{\bf F}$. Then it is easy to
see that ${\bf F}$ $\cong$ ${\bf Z_n}$ (generated by $\alpha$)
and ${\bf {\Gamma}^{\prime}}$ $\cong$ ${\bf {Z_2}{\times}{Z_2}}$
(generated by $\beta {\bf Z_n}$ and $\gamma {\bf Z_n}$).
${\bf {\Gamma}^{\prime}}$ acts freely on ${\bf {C^2}{\times}{T^3}}$;
therefore the singular set of the orbifold ${\bf J}$ consists of
$\{0\} {\times} {\bf T^3/{{\Gamma}^{\prime}}}$, where $\{0\}$
is the origin in ${\bf C^2}$. The singular set is thus a
${\bf {T^3}/{{\Gamma}^{\pr}}}$ family of $A_{n-1}$ singularities.

There is one subtlety pertaining to these examples: there is a non-trivial
monodromy of the singularity for $n$ $\geq$ $2$. This is due to the group
relation
\be
\gamma {\alpha}^k {\gamma}^{-1} = {\alpha}^{-k}
\ee

More generally, if we replace ${\bf Z_n}$ in the above with
any of the finite {\sf A-D-E} subgroups of $SU(2)$, but keep
the group ${\bf {Z_2}{\times}{Z_2}}$ fixed, we obtain a space
${\bf J^G_{{Z_2}{\times}{Z_2}}}$ whose singularities are
a family of {\sf A-D-E} singularities fibered over 
${\bf M_{{Z_2}{\times}{Z_2}}}$. For the cases corresponding to
${\sf D}$ and ${\sf E}$ gauge groups one can in fact check that there
is no monodromy corresponding to outer automorphisms of the
Lie algebra of the gauge group.

Our second example orbifolds are obtained by taking
${\bf K}$ $\cong$ ${\bf {Z_2}{\ltimes}{Z_4}}$. This group is defined
to act in the following way on ${\bf {C^2}{\times}{T^3}}$:

\ba
{\bf Z_2}({\beta}^{\prime}):  (z_1,z_2,x_1,x_2,x_3) &=&
(-z_1,z_2,-x_1,-x_2 + {3 \over 4},x_3 + {1 \over 2})\\
{\bf Z_4}(\eta): (z_1,z_2,x_1,x_2,x_3) &=&
(-iz_2^*,iz_1^*,x_1 + {1 \over 4},-x_2 + {1 \over 4},-x_3)\nonumber
\ea

The orbifolds ${\bf J^{G}_{{Z_2}{\ltimes}{Z_4}}}$ are defined as above
for ${\bf G}$ any ${\sf ADE}$ subgroup of $SU(2)$.
In this case one may check that ${\bf \Gamma}$ $\cong$ ${\bf {G}{\times}
({{Z_2}{\ltimes}{Z_4}})}$. The singular set of  ${\bf J^{G}_{{Z_2}{\ltimes}{Z_4}}}$
consists of a family of ${\sf ADE}$ singularities fibered over
${\bf {T^3}/({{Z_2}{\ltimes}{Z_4}})}$. In these examples, because ${\bf {{Z_2}{\ltimes}{Z_4}}}$
commutes with ${\bf G}$, there is no monodromy of the singularity. The cases in which there is
a non-trivial monodromy will be
discussed in detail in a forthcoming companion paper, since we do not
wish to cloud the main points of this article with these issues.

\bigskip
\noindent
\Large
{\bf {\sf  3. Relation to D6-branes.}}
\normalsize
\bigskip

In this section we will relate $M$ theory on ${\bf J^{Z_n}_{K}}$
to a certain configuration of partially wrapped D6-branes in Type IIA
theory. The purpose of this section is really to confirm the
computation of the massless spectrum of $M$ theory fields supported
near the singularities of ${\bf J}$ and give a clearer origin for
the twisting of the 7-dimensional gauge theory that we discussed
previously. As such this paper can be read independently of this
section.

$M$ theory on ${\bf {C^2}/{Z_n}{\times}R^7}$ gives rise to a
supersymmetric $SU(n)$ Yang-Mills theory which is supported
near the $A_{n-1}$ singularity; hence this gauge theory
lives naturally  on ${\bf R^7}$. In fact, the same
7-dimensional gauge theory arises as the low energy
world-volume theory
on $n$ parallel D6-branes in Type IIA string theory on
${\bf R^{10}}$. The moduli space of this 7-dimensional
super Yang-Mills theory is ${\bf Sym^{n-1}}({\bf R^3})$.
From the point of view of the D6-branes, this moduli space is
easily understood; since they are parallel they all live
at a point in ${\bf R^3}$ and hence we have a ``splitting''
of ${\bf R^{10}}$ $\cong$ ${\bf {R^3}{\times}  R^7}$
corresponding to transverse versus  world-volume dimensions.
Then ${\bf {Sym^{n-1}}(R^3)}$ is just the moduli space of
$n$ particles on ${\bf R^3}$ subjected to one constraint
which eliminates their center of mass motion.
From the point of view of $M$ theory on ${\bf {C^2}/{Z_n}}$,
it is more difficult to understand the gauge theory moduli space
since it is related to the moduli space of hyperkahler structures
on certain ALE-spaces and their orbifold degenerations.
What will be important
is the fact that the $SO(3)$ R-symmetry group of the
gauge theory acts on the coordinates of the various ${\bf R^3}$
factors in the moduli space according to the 3-dimensional 
representation. This $SO(3)$ also acts on the hyperkahler orbifold
${\bf {C^2}/{Z_n}}$ via rotations of the three Kahler forms on this
space. 

Since $M$ theory on ${\bf {C^2}/{Z_n}}$ gives rise to
a gauge theory equivalent to that which comes from $n$ D6-branes
at a point in ${\bf R^3}$, $M$ theory on ${\bf {C^2}/{Z_n}{\times}T^3}$
corresponds from the point of view of the D6-branes to
world-volumes which are no longer ${\bf R^7}$ but rather 
${\bf {T^3}{\times}{R^4}}$. This ``compactification'' of the 7-dimensional
D6-brane world-volume does not affect the ${\bf R^3}$ to which they
are transverse. If we take the 7-dimensional gauge coupling constant
to zero, whilst at the same time shrinking the volume of the 3-torus,
then we can obtain an effective 4-dimensional super Yang-Mills
theory which lives on ${\bf R^4}$. This theory has 4-dimensional
${\cal N}$ = 4 supersymmetry.  
When we further orbifold $M$ theory on 
${\bf {C^2}/{Z_n}{\times}{T^3}}$ by ${\bf K}$ to
give $M$ theory on ${\bf J^{Z_n}_K}$, a natural proposal is that this
procedure commutes with the duality between $M$ theory on 
${\bf {{C^2}/{Z_n}}{\times}{T^3}}$ and the configuration of
$n$ (wrapped) D6-branes in Type IIA theory. Hence,
if this is the case, we expect a dual description of $M$ theory
on ${\bf J^{Z_n}_K}$ in terms of Type IIA theory on 
${\bf ({R^3}{\times}{T^3})/{K^*}}$ with $n$ D6-branes
wrapped around the ${\bf T^3}$ in the cover. Here ${\bf K^*}$
is the image
of ${\bf K}$ under the map from $M$ theory to
Type IIA theory. In particular, since ${\bf K}$ acts freely, the
expectation that the Type IIA dual exists is fairly well
justified \cite{senorb}.

Our next step will be to determine ${\bf K^*}$,
and hence make a proposal for a Type IIA description of $M$ theory on
${\bf J^{Z_n}_K}$. 

The identification of how ${\bf K^*}$ acts
on our wrapped D6-brane configuration
in Type IIA theory on ${\bf {R^3}{\times}{T^3}}$ follows from the simple
observations: $(i)$ the $SO(3)$ {\sf R}-symmetry of the 7-dimensional
super Yang-Mills theory acts via the 3-dimensional representation
on the coordinates of ${\bf R^3}$; $(ii)$ 
in $M$ theory on ${\bf J^{Z_n}_1}$
it acts on the triplet of Kahler forms ${\omega}_i$ living
on ${\bf {C^2}/{Z_n}}$; $(iii)$ in $M$ theory on ${\bf J^{Z_n}_1}$
${\bf K}$ also acts (as a subgroup of $SO(3)$)
on the triplet of 1-forms $dx_i$ in the same
way as it does on the Kahler forms; $(iv)$ since the ${\bf T^3}$
is `common' to both the Type IIA and $M$ theory backgrounds, this
shows that ${\bf K^*}$ acts on the
the coordinates of ${\bf T^3}$ in the same way as ${\bf K}$.
Combining these facts we learn that if
the ${\bf R^3}$ and ${\bf T^3}$ have coordinates
${v_i}$ and ${w_i}$ $\cong$ ${w_i} + 1$ respectively, then
the 1-forms $dv_i$ and $dw_i$ transform under ${\bf K^*}$
in the 3-dimensional representation. Moreover the coordinates
$w_i$ transform under ${\bf K^*}$ in the same way as under
${\bf K}$.

It will be useful
to denote the complex coordinates $Y_i$ as
\be
Y_j \equiv v_j + iw_j
\ee

We can define a Calabi-Yau structure on ${\bf {R^3}{\times}{T^3}}$
by defining 
the Kahler 2-form and holomorphic 3-form
\ba
\omega &=& {1 \over 2}{\delta}^{ij}{dY_i}{\wedge}{dY_j^*}\\ 
\Omega &=& 
{1 \over 6}{\epsilon}^{ijk}{dY_i}{\wedge}{dY_j}{\wedge}{dY_k}
\ea

These forms are clearly preserved by the ${\bf K^*}$ action
on ${\bf {R^3}{\times}{T^3}}$. In fact, they are manifestly 
$SO(3)$ invariant.
Hence, the quotient space
${\bf Y_K}$ $\equiv$ ${\bf ({R^3}{\times}{T^3})/{K^*}}$
admits a Calabi-Yau structure. Furthermore, since ${\bf K^*}$
acts freely on the ${\bf T^3}$ factor, ${\bf Y_K}$ is a smooth complex
threefold, which (although its holonomy is not precisely $SU(3)$)
we can regard as a Calabi-Yau threefold.

We are thus considering Type IIA string theory on
${\bf {Y}{\times}{R^4}}$. This spacetime background preserves
eight supercharges.
We would
now like to re-introduce our
$n$ parallel D6-branes which fill ${\bf R^4}$.
Since the D6-branes world-volume is 7-dimensional, it must
be of the form ${\bf {M}{\times}{R^4}}$, where ${\bf M}$ is
a real 3-dimensional submanifold of ${\bf Y_K}$.
Furthermore, these D6-branes should be invariant under half of
the eight supercharges preserved by the background, since we are
ultimately interested in a configuration with ${\cal N}$ $=$ $1$
supersymmetry.
This will be
the case if, and only if, ${\bf M}$ is a special Lagrangian
submanifold of ${\bf Y_K}$. In fact, as we have discussed above,
the D6-branes are sitting at the origin in the ${\bf R^3}$
and are wrapped around the ${\bf \{0\}{\times}T^3}$ in the 
${\bf {R^3}{\times}{T^3}}$ covering space of ${\bf Y_K}$.
The image of this particular ${\bf T^3}$ submanifold in
${\bf Y_K}$ is simply ${\bf (\{0\}{\times}{T^3})/{K^*}}$
which we thus take as ${\bf M}$. This 
real 3-manifold is indeed a special Lagrangian submanifold
of ${\bf Y_K}$. 
Thus, we propose that the Type IIA configuration
which is dual to $M$ theory on ${\bf J^{Z_n}_K}$ is given by
Type IIA string theory on ${\bf Y_K}$ with $n$ D6-branes
wrapped around the submanifold ${\bf M}$.
 
Let us now consider the Yang-Mills theory which describes the
low energy world-volume dynamics of the D6-branes. We are thus considering
7-dimensional $SU(n)$ super Yang-Mills theory on 
${\bf {M}{\times}{R^4}}$.
Ordinarily, one would conclude that such a theory is not supersymmetric
since ${\bf M}$ does {\it not} admit parallel spinors. However,
${\bf M}$ is a special Lagrangian
submanifold in the Calabi-Yau threefold ${\bf Y}$. and this fact implies
that the super Yang-Mills theory is `partially twisted' and indeed
supersymmetric \cite{dtop}. 
This will be explicitly illustrated shortly. 
Ultimately we are interested in the massless fields which propagate
on ${\bf R^4}$ and so we will think of this 7-dimensional Yang-Mills
theory in a Kaluza-Klein fashion as being compactified from  seven to
four dimensions on ${\bf M}$.

In flat ${\bf R^7}$, the symmetries of the Yang-Mills theory include
\newline
$G$ $\equiv$ 
$Spin(3){\times}Spin(6,1)$. The $Spin(3)$ is the R-symmetry group and
the
$Spin(6,1)$ the double cover of the
spacetime Lorentz symmetry. 
The fields are all massless
and consist of three real scalars ${\phi}_i$ which transforming as
${\bf (3,1)}$ under $G$, the gauge fields $A_{\mu}$, which transform as
${\bf (1,7)}$ under $G$, and two spacetime fermions transforming as
${\bf (2,8)}$. All of these fields transform in the adjoint representation
of the gauge group $SU(n)$.
Compactification of this theory on ${\bf M}$ to
${\bf R^4}$ breaks $G$ to $G^{\pr}$ $\equiv$ 
$Spin(3){\times}Spin(3)^{\pr}{\times}{SL(2,{\bf C})}$, 
where $SO(3)^{\pr}$ $\equiv$ $Spin(3)/{\bf Z_2}$ acts
on the tangent bundle of ${\bf M}$ and 
$SL(2,{\bf C})$ acts on ${\bf R^4}$ as the double cover of the
Lorentz group.
Under $G^{\pr}$, the scalars transform as ${\bf (3,1,1)}$, the
gauge fields split into a gauge field on ${\bf R^4}$ transforming
as ${\bf (1,1,2{\otimes}{\bar 2})}$ 
and three `new' scalars
${\phi}_i^{\pr}$ transforming
as ${\bf (1,3,1)}$. The fermions transform as 
${\bf (2,2,2) + (2,2,{\bar 2})}$. As we remarked above, the lack of
parallel spinors on ${\bf M}$ usually implies that the theory
is not supersymmetric.

Now we must `remember' that we are really discussing D6-branes
wrapped around a supersymmetric cycle in a Calabi-Yau threefold.
From this point of view, the scalars ${\phi}_i$ are 
re-interpreted as sections of the normal bundle $N({\bf M})$
to ${\bf M}$ inside
${\bf Y_K}$ and  the scalars ${\phi}_i^{\pr}$ as sections of the
tangent bundle $T({\bf M})$
to ${\bf M}$. The fermion fields are sections of the
spin bundle of ${\bf M}$ with values in the spin bundle constructed
from the normal bundle. Thus far, we have not utilised the
Calabi-Yau structure of ${\bf Y_K}$. A remarkable
observation due to Mclean \cite{Mcl} asserts that $T({\bf M})$
$\cong$ $N({\bf M})$. In other words, the Calabi-Yau structure
of ${\bf Y_K}$ allows one to identify the normal bundle with the
tangent bundle. 

For our gauge theory, this is the `origin' of
the twisting which implies the theory is supersymmetric.
Since the tangent and normal bundles to ${\bf M}$ have been identified,
{\it both} sets of scalars can be considered as sections
of $T({\bf M})$. On the other hand, if $S({\bf M})$ denotes
the spin bundle of ${\bf M}$, the fermions can now be regarded
as sections of $S{\otimes}S$ $\cong$ 
$\Lambda^0{({\bf M})}{\oplus}\Lambda^1{({\bf M})}$
$\cong$ $\Lambda^0{({\bf M})} {\oplus}T({\bf M})$
In effect, all fields now transform under a group
$G^{\pr\pr}$
$\equiv$ $SO(3)^{\pr\pr}{\times}{SL(2,{\bf C})}$, 
where $SO(3)^{\pr\pr}$
is again the
structure group acting on the tangent
space to ${\bf M}$.
Under $G^{\pr\pr}$ {\it both} sets of scalars
transform as ${\bf (3,1)}$, the gauge fields as 
${\bf (1,2{\otimes}{\bar 2})}$. The fermions now transform as 
${\bf (1,2) + (1,{\bar 2}) + (3,2) + (3,{\bar 2})}$. These 
latter fields are thus two spacetime spinors which are also
scalars on ${\bf M}$ and two spacetime spinors which are
1-forms on ${\bf M}$.

We can finally compute the massless field content of the
gauge theory on ${\bf R^4}$. Fields on ${\bf R^4}$ will be massless
if they are also  zero modes of the derivative operator which acts
on them. The two four dimensional scalars are 1-forms on ${\bf M}$,
so they will give rise to massless fields if the 
(free part of the) first cohomology
group of ${\bf M}$ is non-trivial. Since we have two such
scalars they will contribute $2b_1$ scalars to the four dimensional
theory. The four dimensional gauge fields are zero-forms on
${\bf M}$ so, since $b_o$ is 1, it gives rise to one massless
four dimensional gauge field. The spinors split into zero-forms
and 1-forms on ${\bf M}$, and
they contribute $(1+{b_1})$ massless fermions to the theory.
Also, due to the fact that the compactification on ${\bf M}$
preserves the $SU(n)$ gauge symmetry, all these fields take
values in the Lie algebra of $SU(n)$.
The four dimensional massless field content therefore consists
of one gauge field, $2b_1$ scalars and $1 + b_1$
fermions 
which all transform in the adjoint of the $SU(n)$
gauge group. 
How many supersymmetries exist in the
four dimensional theory ?  The supersymmetries of the theory
are in one to one correspondence with the fermion zero modes.
These number $1 + b_1$. 

Since the first Betti number of ${\bf M}$ is
zero, we find that the massless field content is precisely
that of ${\cal N}$ = 1 super Yang-Mills theory with gauge
group $SU(n)$.

Note that classically ${\cal N}$ $=$ $1$ super Yang-Mills
theory has a $U(1)$ {\sf R}-symmetry group. $Spin(3)$ symmetries
are nowhere to be seen in that theory. How is this fact reconciled
with the statement of the previous paragraph? Firstly,
the supersymmetries and massless fields are all singlets under
the $Spin(3)$ action, so in the theory of interest, this group
acts {\it trivially}. Secondly, the $U(1)$ {\sf R}-symmetry of the 
theory
is also present - it is simply not obvious in the way we
have described the theory. This can easily been seen by noting that
an {\sf R}-symmetry acts on the superspace coordinates. In fact the
superspace measure ${d^2}{\theta}$ transforms under $U(1)$
{\sf R}-symmetries with a charge minus that of the holomorphic 3-form
$\Omega$. In fact, if we canonically normalise so that under
a $U(1)$ rotation ${d^2}{\theta}$ transforms with charge $-2$,
then the $U(1)$ {\sf R}-symmetry of the model acts on the coordinates of
${\bf Y}$ as follows:
\be
U(1): Y_j \rightarrow e^{{2 \over 3}i{\alpha}} Y_j
\ee

This is clearly not a manifest symmetry of the wrapped
D6-brane configuration. Therefore, the symmetries of the
Yang-Mills theory are not obviously in contradication
to those of the D6-branes wrapped on ${\bf M}$. 
Suppose, for arguments sake, that the D6-branes were wrapped
around a special Lagrangian submanifold ${\bf M^{\pr}}$ in some
other Calabi-Yau threefold ${\bf Y^{\pr}}$. Further suppose that 
${b_1}({\bf M^{\pr}})$ $=$ $1$. Then, according to the above
analysis, the effective four dimensional theory has ${\cal N}$
$=$ $2$ supersymmetry. Classically, a field theory with these
supersymmetries has an $Spin(3){\times}U(1)$ {\sf R}-symmetry group. 
However, in the D6-brane model, only the $Spin(3)$ symmetry 
is obvious.

\newpage
\noindent
\large
{\bf {\sf 4. N = 1 Super Yang-Mills.}}
\normalsize
\bigskip

For completeness and in order to compare easily with $M$ theory
results obtained later we briefly give a review of ${\cal N}$
$=$ $1$ pure super Yang-Mills theory. More detailed reviews
can be found in \cite{SYM}. We begin with gauge group $SU(n)$.
${\cal N}$ $=$ $1$ $SU(n)$ super Yang-Mills theory
in four 
dimensions is an extensively studied quantum field
theory. The classical Lagrangian for the theory is
\be
{\cal L} = -{1 \over 4{g^2}}(F_{\mu\nu}^{a})^2 +
{1 \over {g^2}}{\bar \lambda}^{a}i{\not{\cal D}}{\lambda}^a +
i{\theta \over 32{\pi}^2}F_{\mu\nu}^{a}{\tilde F}^{a\mu\nu}
\ee
$F$ is the gauge field strength and $\lambda$ is the gaugino field.

It is widely believed that
this theory exhibits dynamics very similar to
those of ordinary QCD: confinement, chiral symmetry breaking,
a mass gap. Supersymmetry constrains the dynamics of the
theory so strongly, that the low energy  
effective superpotential for the model is known. 
It is given by
\be
W_{eff} = c{\mu}^{3}e^{2{\pi}i{\tau}/{n}}
\ee
here $\tau$ is the complex coupling constant,
\be
\tau = {\theta \over 2\pi} + i{4\pi \over g^2}
\ee
and $\mu$ the mass scale.

In particular, the form of this potential suggests that it is
generated by dynamics associated with ``fractional instantons'', ie
states in the theory whose quantum numbers are formally of
instanton number ${1 \over n}$. Such states are also closely
related to the spontaneously broken chiral symmetry of the theory.
Let us briefly also review some of these issues here.

Under the $U(1)$ {\sf R}-symmetry of the theory, the gauginos transform
as
\be
\lambda \rightarrow e^{i\alpha}\lambda
\ee
This is a symmetry of the classical action but not of the quantum
theory (as can easily be seen by considering the transformation
of the fermion determinant in the path integral). However, if the above
transformation is {\it combined} with a shift in the theta angle
of the form
\be
\tau \rightarrow \tau + {2n \over 2\pi}\alpha
\ee
then this cancels the change in the path integral measure. 
This shift symmetry is a bona fide symmetry of the physics
if $\alpha$ $=$ ${2\pi \over 2n}$, so that even in the
quantum theory a ${\bf Z_{2n}}$ symmetry remains. Associated
with this symmetry is the presence of a non-zero value for the
following correlation function,
\be
\langle\lambda\lambda(x_1)\lambda\lambda(x_2)...
\lambda\lambda(x_n)\rangle
\ee
which is clearly invariant under the ${\bf Z_{2n}}$ symmetry.
This correlation function is generated in the 1-instanton
sector and the fact that $2n$ gauginos enter is due to the
fact that an instanton of charge 1 generates $2n$ chiral
fermion zero modes.

Cluster decomposition implies that the above correlation function
decomposes into `$n$ constituents' and therefore there exists a
non-zero value for the gaugino condensate:
\be
\langle \lambda \lambda \rangle {\neq} 0
\ee

Such a non-zero expectation value is only invariant under a
${\bf Z_2}$ subgroup of ${\bf Z_{2n}}$ implying that the
discrete chiral symmetry has been spontaneously broken.
Consequently this implies the existence of $n$ vacua in the theory.

In fact, it can be shown that
\be
\langle\lambda\lambda\rangle = 16{\pi}i{\partial \over \partial\tau}
W_{eff} = -{32{\pi}^2 \over n}c{\mu}^{3}e^{2{\pi}i{\tau}/n}
\ee

In view of the above facts it is certainly tempting to propose
that `fractional instantons' generate the non-zero gaugino
condensate $(18)$ directly. But this is difficult to see
directly in SQCD. We will see that this is precisely what 
the dynamics of $M$ theory on ${\bf J^{z_n}_K}$ predicts.

More generally, if we replace the $SU(n)$ gauge group by
some other gauge group $H$, then the above statements are
also correct but with $n$ replaced {\it everywhere}
with ${c_2}(H)$ the 
dual Coxeter number of $H$. 
For {\sf A-D-E} gauge groups
${c_2}(H)$ $=$ ${\Sigma}_{i=1}^{r+1} a_i$, where $r$ is the
rank of the gauge group and the 
$a_i$ are
the Dynkin indices of the affine Dynkin diagram associated
to $H$. For $A_n$, all the $a_i$ $=$ $1$; for $D_n$ groups
the four `outer' nodes have index $1$ whilst the rest have $a_i$ 
$=$ $2$. $E_6$ has indices $(1,1,1,2,2,2,3)$, $E_7$
has $(1,1,2,2,2,3,3,4)$ whilst $E_8$ has indices
$(1,2,2,3,3,4,4,5,6)$.

\bigskip
\noindent
\large
{\sf 5. Theta angle and Coupling Constant in M theory.}
\normalsize
\bigskip

The physics of $M$ theory supported near the singularities of
${\bf {C^2}/{G}{\times}{R^7}}$ is described by
super Yang-Mills theory on ${\bf R^7}$. The gauge coupling constant
of the theory is given by
\be
{1 \over g^2_{7d}} \sim {1 \over l^3_p}
\ee
where $l_p$ is the eleven dimensional Planck length. In seven dimensions,
the analog of the theta angle in four dimensions is actually a
three-form $\Theta$. The reason for this is the seven dimensional
interaction
\be
L_I \sim \Theta {\wedge}F{\wedge}F
\ee

(with $F$ the Yang-Mills field strength). In $M$ theory $\Theta$
is given by $A$, the three-form potential for the theory. 

If we now take $M$ theory on ${\bf J^G_K}$ we have essentially
`compactified the seven dimensional theory with a twist' and the
four dimensional gauge coupling constant is roughly given by
\be
{1 \over g^2_{4d}} \sim {V_M  \over l^3_p}
\ee
where $V_M$ is the volume of ${\bf M}$.
The four dimensional theta angle can be identified as
\be
\theta = \int_{\bf M} A
\ee
The above equation is correct because under a 
global gauge transformation
of $A$ which shifts the above period by $2\pi$ times
an integer - a transformation which is a symmetry of $M$
theory - $\theta$ changes
by $2\pi$ times an integer. Such shifts in the theta angle
are also global symmetries of the field theory. 

Thus the complex gauge coupling constant of the effective four
dimensional theory may be identified as
\be
\tau = \int_{\bf M} ({A \over 2\pi} + ic{{\varphi} \over l^3_p})
\ee
($c$ is a normalisation constant)   
The quantities on the right hand side in the above expression
are the real and imaginary parts of a complex scalar field
in the theory which is the bosonic part of the {\it single}
massless chiral superfield in the background ${\bf J^G_K}$.
(The reason that there is only one massless chiral superfield
stems from the fact that $b_3^{str}$, the third string theoretic
Betti number of ${\bf J^G_K}$ is equal to one.)
In view of this we should really view $\tau$ as a background
superfield and regard the above equation as an expression
for its vacuum expectation value.

\bigskip
\noindent
\Large
{\bf {\sf 6. BPS States.}}
\normalsize
\bigskip

In any ${\cal N}$ $=$ $1$ supersymmetric theory,
a natural class of states to discuss are the BPS saturated
states. 
These are states which are invariant under {\it two}
of the four supercharges. In general, such a theory
can contain BPS instantons, strings and domain walls.
However, in the pure ${\cal N}$ $=$ $1$ super Yang-Mills
theory which interests us, only BPS instantons and domain
walls can appear. A simple and natural question to pose is
where are these states in the $M$ theory and Type IIA
backgrounds we have been discussing? 

The instantons will be of primary interest to us
in this paper.

\bigskip
\noindent
{\sf 6.1 M2-brane Instantons.}
\normalsize
\bigskip

In $M$ theory on ${\bf {C^2}/{G}{\times}{R^7}}$, the $M$ theory state
which corresponds to an instanton of charge {\it one} in
the super Yang-Mills theory is {\it one} $M$2-brane whose world-volume
is of the form ${\bf R^3}$ $\subset$ ${\bf R^7}$. We can
think of this charge one $M$2-brane as $|{\bf G}|$
${\bf G}$-equivariant $M$2-branes on ${\bf {C^2}{\times}{R^7}}$.
However, because  ${\bf G}$ does not act on the ${\bf R^4}$
which is transverse to the world-volume of the $M$2-branes in spacetime,
these $|{\bf G}|$ $M$2-branes can be at different
positions in spacetime. From the point of view of a 
low energy field theory observer in ${\bf R^7}$, each of these
branes carry fractional instanton number. These are just the fractional
branes described in \cite{frac}.

The
precise amount of charge carried by these branes
is naturally encoded in the 
extended Dynkin diagram associated with the {\sf A-D-E}
singularity. As explained in \cite{dougmoore}, the dynamics
of a charge one D-brane moving on ${\bf {C^2}/{G}}$ is described
by a quiver gauge theory on the world-volume
whose quiver diagram is related to the
affine Dynkin diagram of the {\sf A-D-E} group.
Let $a_i$ denote the Dynkin index for the $i$'th
node of such a diagram; then the gauge group of the
theory is ${\Pi}_{i} U(a_i)$ which means it is
natural to associate $a_i$ D-branes to the $i$'th node.
Since
\be
{\Sigma}_{i=1}^{r+1} a_i = {c_2}(H)
\ee
where $r$ is the rank of the group $H$,
the single D-brane `fractionates'
into $r$ + $1$
fractional D-branes the $i$'th of 
which carries charge ${a_i \over {c_2}(H)}$.
In the $M$ theory lift of this structure that we wish
to consider here, with the D-brane replaced by the $M$2-brane,
it is natural to assume that the same fractionation of charge
takes place.

For example, in the case when ${\bf G}$ is ${\bf Z_n}$ so that
the gauge group is $SU(n)$, all the $n$ nodes have index $1$
and the single $M$2-brane instanton fractionates into $n$ identical
fractional instantons each of charge ${1 \over n}$.

Now consider the case when we geometrically engineer using
${\bf J^G_K}$. The $M$ theory states which correspond to field theory
instantons now correspond to $M$2-branes wrapped around 
the supersymmetric 3-cycle ${\bf M_K}$. However, the group
${\bf G}$ is still acting transverse to these $M$2-branes and
hence one can see that the BPS spectrum of $M$ theory on ${\bf J^G_K}$
contains fractional $M$2-brane instantons.

We have argued that $M$ theory on ${\bf {J^G_K}{\times}{R^4}}$ gives
rise to a geometrically engineered version of  super Yang-Mills
theory with four supercharges and
{\sf A-D-E} gauge group. 
Furthermore, we have identified 
instantons of the field theory as $M$2-branes in $M$ theory.
Since in the super Yang-Mills theory
in question, a charge {\it one} instanton generates $2{c_2}(H)$ 
chiral fermionic zero modes, we can fairly
safely {\it assume} that the corresponding charge one
wrapped $M$2-brane also generates precisely $2{c_2}(H)$ 
such zero modes.
This has the following simple consequence:
a fractional $M$2-brane instanton of charge ${1 \over {c_2}(H)}$
generates {\it two} fermionic zero
modes. This implies that they can in principle generate a superpotential
in the four dimensional theory.  If such a superpotential is generated,
it will be of the form,
\be
W = C.e^{-S_I}
\ee
where $S_I$ is the supersymmetric 
action for the $M$2-brane instanton
(and $C$ is a dimensionful normalisation constant). $S_I$ will
contain a contribution proportional to the volume of the $M$2-brane
world-volume (ie ${\bf M}$), with a constant of proportionality
determined by the topological charge.

In fact
\be
W = C.e^{{2\pi i\tau}/{{c_2}(H)}}
\ee

The reason that the exponent comes precisely with the
above factor is that $exp(-S_I)$
for an instanton of
charge {\it one} must be given by
\be
e^{2\pi i\tau}
\ee
This is because the real part of the exponent
is minus the action for {\it one} instanton and 
the  superpotential
must be holomorphic.
A convenient
way of writing $(27)$ is
\be
C^{\pr}.e^{2\pi i\tau} = {\Pi}_{i=1}^{r+1} W^{a_i}
\ee
Hence, we can interpret $W^{a_i}$ as $exp(-S^{a_i}_I)$,
where $S^{a_i}_I$ is the action for a fractional instanton
of charge ${a_i \over {c_2}(H)}$.
Thus we see that equation $(26)$ is just the action for
those fractional instantons associated with nodes
of index $1$ (ie with topological charge ${1 \over {c_2}(H)}$).

Clearly the result $(26)$ agrees with the field theory
expectation $(12)$. Moreover the fact that the superpotential is
dynamically generated by an object carrying a fractional instanton
number also agrees with the results of \cite{gom1,gom2,brodie,zab}.

\bigskip
\noindent
{\sf 6.2 Relation to Membrane Anomalies.}
\bigskip

In field theory instantons are usually associated with
anomalies. Whether or not
the $M$5-brane instantons considered in \cite{Wit}
generated a superpotential was shown to be related to
a certain anomaly in rotations of the normal bundle to
the $M$5-brane. On the other hand, $M$2-branes have
odd dimensional world-volumes and the only known anomaly
associated with them is the so-called parity anomaly on
the world-volume \cite{manom}. This can occur if the world-volume
of the $M$2-brane is transported around a transverse circle.

For a charge {\it one} $M$2-brane with world-volume ${\bf M}$,
the path integral for $M$ theory includes the factor
\be
e^{i\int_{\bf M} A}
\ee

The change in this factor as the world-volume is transported
around a transverse circle ${\bf S^1}$ is
\be
e^{i{\int_{\bf {M}{\times}{S^1}} (\lambda \pi + G)}}
\ee
where $\lambda$ is a certain characteristic class \cite{manom}
and $G$ is the
field strength of $A$.
Since $2\pi Re\tau$ $=$ $\int_{\bf M} A$, from the point of
view of our geometrically engineered theory, the above
change can be interpreted as a shift in the theta angle
\be
\tau \rightarrow \tau + 
{{\int_{\bf {M}{\times}{S^1}} (\lambda \pi + G)} \over 2\pi}
\ee
 
If we compare to field theory, then (at least for certain
choices of circles around which the membrane is transported)
it is natural to try and identify this rotation as
a consequence of
an {\sf R}-symmetry transformation. 
If we let

\be
e^{i{\int_{\bf {M}{\times}{S^1}}} (\lambda\pi + G)} = e^{ik\alpha}
\ee

we wish to determine the `charge' $k$.
We can determine $k$ as follows.
In field theory, if we normalise the {\sf R}-symmetry generator
as in $(14)$ so that the gauginos transform with charge one,
then the superpotential transforms with charge two. 
Equation $(28)$ then implies that

\be
k = 2{\Sigma}_i a_i = 2{c_2}(H)
\ee
With the charge determined we see that equation $(31)$
becomes
\be
\tau \rightarrow \tau + {2{c_2}(H) \over 2\pi}\alpha
\ee
This agrees precisely with the field theory result
$(15)$. Hence if our assumptions about the {\sf R}-symmetry are
correct $M$ theory ``knows'' that the chiral symmetry is
${\bf Z_{2{c_2}(H)}}$ since the cancellation of the
membrane anomalies implies that $(34)$ is a symmetry.
The generation of the superpotential
$(26)$ then implies that this chiral symmetry is
spontaneously broken to ${\bf Z_2}$ by fractional
$M$2-brane instantons. This is a rather satisfying
result. 

\newpage
\noindent
\large
{\bf {\sf Acknowledgements.}}
\normalsize
\bigskip

The author would like to thank Steve Thomas for comments,
E. Witten for discussions in 1996 concerning 
membrane instanton generated superpotentials and G. Moore
for informing us about \cite{harvmoore} prior to
publication. We would also like to acknowledge discussions with
D. Morrison and R. Plesser during which we realised the existence of
non-trivial monodromies in certain examples.
The author is supported by a PPARC Postdoctoral Fellowship.

\end{document}